\documentclass[10pt]{article}
\usepackage[super,compress]{cite}\usepackage{url}

\usepackage{amsmath,amsfonts,amssymb,bm,slashed,bbm}
\usepackage{graphicx,color}
\renewcommand{\Ref}[1]{\eqref{#1}}
\newcommand{\eq}[2]{\begin{align}\label{#1}#2\end{align}}

\newcommand{\nn}{\nonumber}
\renewcommand{\ni}{\noindent}
\newcommand{\pa}{\partial}
\newcommand{\ep}{\epsilon}
\newcommand{\ka}{\kappa}
\newcommand{\al}{\alpha}
\newcommand{\ga}{\gamma}
\newcommand{\la}{\lambda}\newcommand{\om}{\omega}

\renewcommand{\L}{{\cal L}}
\newcommand{\rGP}{GP~}

\begin{document}

\title{Casimir effect with an unstable mode}

\author{
		M. Bordag\footnote{bordag@mail.ru}\\ 
\small 	Bogoljubov Laboratory of Theoretical Physics, Joint Institute for Nuclear Research\\
\small	141980 Dubna, Russian Federation\\ 
\small 	Institute for Theoretical Physics, University Leipzig\\
\small	IPF 231101, D-04081 Leipzig, Germany
	 \and 
		 I.G. Pirozhenko\footnote{pirozhen@theor.jinr.ru} \\
\small	 Bogoljubov Laboratory of Theoretical Physics, Joint Institute for Nuclear Research\\
\small	 141980 Dubna, Russian Federation
	 }

 \date{February 13, 2025}
\maketitle

\begin{abstract}
	We consider the Casimir effect in a (1+1)-dimensional model with a critical mode. Such a mode gives rise to a condensate described by the nonlinear Gross-Pitaevskii equation. In the condensate, there are two sources of the Casimir force; one is the conventional one resulting from the fluctuations, the other follows from the condensate. We consider three simple models that allow for condensate solutions in terms of elliptic Jacobi functions. We also investigate a method for obtaining approximate solutions and show its range of applicability. In all three examples we compute the condensate energy. In one example with a finite interval with Robin boundary conditions on one side and Dirichlet conditions on the other side, we calculate the vacuum energy and the Casimir force. There is a competition between the forces from the condensate and the fluctuations. We mention that the force from the condensate is always repulsive.
\end{abstract}


\section{\label{T1}Introduction}	
The Casimir effect, in a narrow sense, is the attraction between two bodies with reflective surfaces, such as parallel plates, in a vacuum. In a broader sense, it is the separation dependence of the 
energy in the presence of boundaries; sometimes the vacuum energy itself is meant.
  The latter can be represented as a half-sum,
\eq{1.1}{ E_0=\frac{\hbar}{2}\sum_J\om_J,
}
over the eigenfrequencies $\om_J$ of the corresponding fluctuation operator, usually some Hamiltonian $H$,
\eq{1.2}{H \phi_J&=\om_J^2\ \phi_J.
}
In a field theory, the sum in \Ref{1.1} is infinite and diverges in the ultraviolet. The most convenient and common regularization uses the zeta function of the operator $H$,
\eq{1.3}{ \zeta_H(s)=\sum_J \om_j^{-2s},~~~(s>\frac12 \mbox{ in one dimension}),
}
as regularization,
\eq{1.4}{ E_0(s)&=\frac{\hbar}{2}\, \zeta_H(s-\frac12),
}
with $s\to0$ at the end. The zeta function is a meromorphic function with an analytic continuation to $s=0$. The treatment of divergences in terms of a heat kernel expansion is well known, see for example the book Ref. \cite{BKMM}.

So far, the topic described above has been considered for stable systems, i.e. for systems with stable modes whose eigenfrequencies $\om_J$ are real. Beyond the Casimir effect, systems with unstable modes are a well-studied topic. In this case, one or more eigenfrequencies have an imaginary part, $\Im \om_j>0$, leading to a decay {of the perturbative vacuum} of the system with the creation of particles. Physically, there is always some mechanism that stops this process. For fermions this will be the Pauli principle, for charged particles the Coulomb repulsion. For {neutral} bosons, you need a repulsive self-interaction, such as in an SU(2) model. In the simplest case you have a real scalar field $\phi(x)$ with a Lagrangian
%
\eq{1.5}{\L &= \int dx\ \left[- \frac12\phi\left(\pa_t^2-\Delta+m^2+V(x)\right)\phi
	-\frac{\la}{4}\phi^4 \right],
}
with the $\phi^4$ self-interaction stabilizing the system. In a linearized version of this theory, the potential $V(x)$ is assumed to be sufficiently attractive to induce a bound state solution,
\eq{1.6}{\left(-\Delta+m^2+V(x)\right)\phi_{bs}=-\ep^2\phi_{bs},
}
so that  we have in the above notations, eqs. \Ref{1.1} and \Ref{1.2},  at least one imaginary frequency,
\eq{1.7}{ \om_{J_0} =i\ep.
}
For the existence of such a bound state solution there is a threshold in the strength of the potential where the binding energy is $\ep=0$ and $\ep> 0$ afterwards.

From an energy point of view, in such a situation the system is at the 'top of the hill' for $\phi=0$. In particular, for a constant potential with $m^2+V<0$, this is the well-known situation of the Higgs mechanism, {which was discussed in Ref. \citen{cole73-7-1888}} and which {became} a cornerstone of the Standard Model. The system will leave the top and go down. Mathematically, this is equivalent to a field shift,
\eq{1.8}{\phi \to\phi_0 {(x)}+\phi{(t,x)}.
}
The field $\phi_0$ (we assume it is time independent) is the condensate and the new $\phi$ describes the quantum fluctuations.
The new Lagrangian,
\eq{1.9}{\L&= \L_0+\L_1+\L_2+\dots,
}
consists of several parts,
\eq{1.10}{\L_0 &= \int dx\ \left[- \frac12\phi_0\left(-\Delta+m^2+V(x)\right)\phi_0
	-\frac{\la}{4}\phi_0^4 \right],
	\\\nn
	\L_1 &= - \int dx\ \phi \left(-\Delta+m^2+V(x)+\la \phi_0^2\right)\phi_0,
	\\\nn
	\L_2 &= - \int dx\ \frac12\phi\left(\pa_t^2-\Delta+m^2+V(x)+3\la\phi_0^2\right)\phi.
}
Here $\L_0$ depends only on the condensate field. Its energy is
\eq{1.11}{E_{cond}=\int dx\ \left[ \frac12(\nabla\phi_0)^2 +m^2+V(x))\phi_0^2
	+\frac{\la}{4}\phi_0^4\right].
}
The linear contribution in $\phi$, $\L_1$, must vanish, which leads to a kind of real version of the Gross-Pitaevskii (\rGP) equation
\eq{1.12}{ \left(-\Delta+m^2+V(x)+\la\phi_0^2\right)\phi_0=0.
}
For a complex field, the GP equation is well known in a number of areas, ranging from Bose-Einstein condensation to pion condensation. In this case it is equivalent to the nonlinear Schr\"odinger equation. The real version, \Ref{1.12}, without mass and potential, describes the classical anharmonic oscillator.  


Taking a solution $\phi_0$ of the \rGP equation, which is generally position dependent, $\phi_0(x)$, the system will be stable after the shift \Ref{1.8}. This means that the spectrum of the operator for the fluctuations, i.e. the kernel in $\L_2$, has only real eigenfrequencies and the corresponding vacuum energy can be calculated by standard methods. It will have no imaginary part. Finally, the dots in \Ref{1.9} indicate self-interaction of the quantum field, which is beyond the scope of this paper.

To the best of the author's knowledge, the Casimir effect has only been considered for stable systems. The present paper aims to fill this gap and we consider some simple examples of systems with one unstable mode for a real scalar field with the Lagrangian \Ref{1.5} in (1+1)-dimensions.  Generalizations to parallel plates in higher dimensions should be easy.

Since the \rGP equation \Ref{1.12} is non-linear,  there are no general methods to solve it and one is left with approximations. However, in the (1+1)-dimensional example considered, there are exact solutions in terms of an elliptic function. In addition, we consider an approximation scheme based on the mean-field approximation and show that it is applicable for weak instability, i.e., for $\ep\gtrsim 0$ in \Ref{1.6}.

We develop a perturbative scheme and also compute the exact solutions in terms of elliptic functions. The perturbative approach amounts to taking the bound state solution, $\phi_{bs}$, as a first approximation to the solution $\phi_0$ of the \rGP equation. We show that this approximation is valid for small binding energies, $\ep$, of the bound state, and indicate how to calculate corrections.

To put this work in the context of quantum field theory, we mention that the condensation of Bose particles is an interesting and important topic, especially in a strong external field or in a high-density medium. {As mentioned above, unlike fermions, which are subject to the Pauli principle, bosons need some mechanism to stop the condensation. This can be the repulsive Coulomb force, for example for charged ions, or the self-repulsion of nonabelian fields. {In strong interaction theory, pions are the relevant bosons. Their} condensation may occur in superheavy nuclei, as first proposed in Ref. \citen{migd72-34-1184}, in neutron stars Ref. \citen{latt10-54-101}, or in heavy ion collisions Ref. \citen{vosk94-78-793}. However, an experimental confirmation is still missing.							
 	
A condensate appeared in connection with the Casimir force in attempts to measure the influence of superconductivity on the force. The problem is as follows. While the transition to the superconducting phase is accompanied by an abrupt change in the dc conductivity, its influence on the Casimir force is only in the far infrared and is small. In Refs. \citen{bimo05-726-441}, \citen{bimo05-94-180402} it was mentioned that the Casimir energy of the order of the condensation energy in the superconductor and a change of the transition temperature or the associated critical magnetic field might be measurable. However, this does not seem to be as easy as expected, Ref. \citen{pere20-6-115}.

Since we are going to use the elliptic Jacobi function as solutions of the \rGP equation, we mention the earlier use of these functions for the wave functions of Bose-Einstein condensates. In Ref. \citen{carr00-62-063610} and Ref. \citen{carr00-62-063611} such a condensate was considered in a box with several boundary conditions. For example, bright and dark solitons were discussed. In Ref. \citen{mahm02-66-063607} a double square well potential was used, and finally in Ref. \citen{raga24-216-814} a delta function potential inside the well was used. In most cases interesting nonlinear effects such as bifurcations were observed.
To some extent related is the study of a nonlinear Schr\"Odinger equation with a delta-function perturbation in Ref. \citen{haki97-55-2835}.
For a particle obeying the \rGP equation in a box, the solutions were studied in Ref. \citen{carr00-15-2645} and the Casimir effect was calculated in Ref. \citen{bord21-7-55}.

As a basic example, we consider a cavity with a Robin boundary condition.
Depending on the parameters, it can provide a bound state similar to the delta function potential. When the binding energy exceeds the mass, such a system has a critical mode. The Casimir effect with Robin boundary conditions was studied more than 20 years ago in Refs. \citen{rome02-35-12973} and \citen{bord02-65-064032}. While in the first paper only the non-critical case was considered, in the second the critical mode was included, but without condensate (and a zeta function with imaginary part was obtained). Also in this paper a connection of the Robin boundary conditions with the brane-world scenario was discussed.

We start with the general formulas in section \ref{T2}.
Then we consider the simplest example, a delta function potential on the whole axis. In section \ref{T4} we consider the case with Robin boundary conditions. Then, as another example, we consider a potential hole. Finally, in the next section we calculate the vacuum energy and the Casimir force for the second example. After that the conclusions are given.

\ni Throughout the paper we use natural units.

 \section{\label{T2}Basic formulas and model setup}
We start with the Lagrangian \Ref{1.5} for a (1+1)-dimensional real field $\phi(t,x)$.
\eq{2.1}{\L &= \int dx\ \left[- \frac12\phi(t,x)\left(\pa_t^2-\pa_x^2+m^2+V(x) \right)\phi(t,x)
	-\frac{\la}{4}\phi(t,x)^4 \right].
}
In the first example, a delta potential, we take
\eq{2.2}{ V(x) &= -2\ka\delta(x),\qquad x\in(-\infty,\infty),
}	
where $\ka>0$ is the parameter describing the depth of the potential. This potential is realized by the matching condition,
\eq{2.3}{\phi'(t,+0)-\phi'(t,-0)=-2\ka\,\phi(t,0),\quad \phi(t,+0)=\phi(t,-0).
}	
As a second example, we consider a field on the finite interval $x\in[0,L]$ with boundary conditions,
\eq{2.4}{ \left(\kappa+\pa_x\right)\phi(t,x)_{|{x=0}}&=0,\qquad \mbox{Robin},
	\\\nn
	\phi(t,x)_{|x=L}&=0,\qquad\mbox{Dirichlet}.
}	
In this case we have no potential ($V=0$). However, we will keep the notation '$V(x)$' to indicate that we're not working with a free particle.
Finally, as a third example, we take a potential hole on the half axis, $x\in[0,\infty)$,
\eq{2.4a}{V[x]=\left\{\begin{array}{rl}-U_0, & 0<x<R, \\ 0,& R<x.\end{array}\right.
}
At $x=0$ the field satisfies the Dirichlet boundary condition, $\phi(0)=0$.

In all cases we assume that the potential is strong enough to have an unstable mode with \Ref{1.7}.

In (1+1) dimensions, the \rGP equation \Ref{1.12} takes the form
\eq{2.4b}{\left( -\pa_x^2+m^2+V(x)+\la\phi_0^2(x) \right)\phi_0(x) =0.
}
Since we have assumed that the potential is independent of time, the condensate function $\phi_0(x)$ is static. In all examples considered, the \rGP equation can be solved exactly, see below. Then the energy \Ref{1.11} of the condensate can be simplified using \Ref{1.12},
\eq{2.5}{ E_{cond}=-\frac{\la}{4}\int dx\, \phi_0(x)^4.
}	
In the first example, the integration is over the entire axis, in the second from $0$ to $L$, and in the third from zero to infinity.

In the approximation scheme we consider a kind of mean-field approximation in the \rGP equation, \Ref{2.4b}, by replacing the nonlinear contribution by a constant, $\la\phi_0^2(x)\to \ep^2$, and arrive at the equation
\eq{2.5a} { \left( -\pa_x^2+m^2+V(x) \right)\phi_{bs}(x) =-\ep^2 \phi_{bs}(x).
}
This equation is formally equivalent to the Euler-Lagrange equation which follows from the Lagrangian $\L$, \Ref{1.5}, for $\la=0$. {In the considered overcritical case it is a bound state equation for the potential $V(x)$ as in quantum mechanics.  It is a linear equation and we know that it has normalizable solutions, $\phi_{bs}(x)$, with an energy level below zero, $-\ep^2$. We mention that it also has bound state solutions with $0<\om^2=-\ep^2<m^2$, but these are not critical.

Using this function as a {\it approximation} to the solution of the \rGP equation \Ref{1.12}, we get the formula
\eq{2.6}{\phi_0(x) &= \mu\phi_{bs}(x)+\delta\phi(x),
}	
where $\mu$ is a normalization parameter that remains free since \Ref{2.5a} is a linear equation.

For $\delta\phi(x)$ we develop a perturbative expansion. For this we insert the ansatz \Ref{2.6} into the \rGP equation \Ref{1.12}, and with \Ref{1.6} we get
\eq{2.7}{ \left(-\pa_x^2+m^2+V(x)\right)\delta\phi(x) &=
	\ep^2\mu^2\phi_{bs}(x)-\la\left(\phi_{bs}(x)+\delta\phi(x)\right)^3.
}	
To continue, we define the Green's function of the operator on the left side,
\eq{2.8}{ \left(-\pa_x^2+m^2+V(x)\right) G(x,x') &=\delta(x-x'),
}	
and rewrite \Ref{2.7} in the form
\eq{2.9}{\delta\phi(x) &= \int dx'\, G(x,x')\left[
	-\ep^2\mu\phi_{bs}(x')-\la\left(\mu\phi_{bs}(x')+\delta\phi(x')\right)^3\right].
}	
This equation can be solved by iteration, starting from $\delta\phi(x)=0$, and we get some expansion for $\delta\phi(x)$.  To find the 'small parameter' of this expansion, we look at the energy. We also need to find the parameter $\mu$, which is arbitrary so far.
This can be done by considering $\mu$ as a variational parameter and requiring that it give a minimum to the energy \Ref{1.11}.

We consider the first step of the iterations and put $\delta\phi(x)=0$ on the right side of \Ref{2.9},
\eq{2.10}{\delta\phi^{(1)}(x) &= \int dx'\, G(x,x')\left[
	-\ep^2\mu\phi_{bs}(x')-\la\mu^3\phi_{bs}(x')^3\right].
}
By inserting $\delta\phi^{(1)}(x) $ into the right side of \Ref{2.9}, we get higher powers of $\ep$ and $\mu$. So we conclude that the iteration gives an expansion in powers of $\ep$ and $\mu$.

Next we look at the energy. Inserting the leading approximation, $\phi_0(x)=\phi_{bs}(x)$, into \Ref{1.11} yields the energy $E_{bs}$ of the bound state in the form
\eq{2.11}{ E_{bs} &= -\frac12 \ep^2\mu^2 a+\frac{\la}{4}\mu^4b,
}	
where
\eq{2.12}{ a&=\int dx\, \phi_{bs}(x)^2,\quad b=\int dx\, \phi_{bs}(x)^4,
}	
depend on the specific model used. This energy can be rewritten in the form
\eq{2.13}{ E_{bs} &= \frac{\la b}{4}\left(\mu^2
	-\frac{\ep^2a}{\la b}\right)^2-\frac{\ep^4a^2}{4\la b}.
}	
The minimum is in
\eq{2.14}{\mu^2=\frac{\ep^2a}{\la \, b},
}	
so if $\ep$ is a small parameter, so is $\mu$.
Inserting this $\mu$ into \Ref{2.10}, we get the first correction in the form of
\eq{2.15}{\delta\phi(x) &= -\ep^2\sqrt{\frac{a}{4b}}
	\int dx'\, G(x,x')\left[
	\phi_{bs}(x')+\frac{\la a}{4b}\phi_{bs}(x')^3\right].
}	
As you can see, $\ep$ is the small parameter for which the approximation \Ref{1.8} makes sense. With \Ref{2.14}, the energy of the bound state is
\eq{2.16}{ E_{bs} &= -\frac{\ep^4 }{4\la }\, \frac{a^2}{b}.
}	
It is an approximation from above to the exact condensate energy \Ref{2.5},
\eq{2.17}{ E_{bs}> E_{exact},
}	
for small $\ep$. We will check this relation in the following examples.

As for the fluctuations, following the Lagrangian $\L_2$ in \Ref{1.10}, we should mention that the corresponding spectrum,
\eq{2.18}{ \left( -\pa_x^2 +m^2+V(x)+3\la\phi_0(x)^2\right)\phi=\om^2\phi,
}
is stable by construction if a solution of the \rGP equation \Ref{1.12} is substituted for $\phi_0$. When using an approximation as suggested in \Ref{2.6}, e.g. a bound state function $\phi_{bs}$, stability is not guaranteed.

\section{\label{T3}Delta Function Potential}
In this section we consider the example with the delta function potential \Ref{2.2}. As we know, the bound state solution is
\eq{3.1}{ \phi_{bs} &= e^{-\ka|x|},\quad \ep=\sqrt{\ka^2-m^2}.
}	
In the critical case we have $\ka>m$. This solution satisfies the equation \Ref{1.6} with $V(x)$, \Ref{2.2}, or equivalently with $V(x)=0$ and the matching condition \Ref{2.3}. The expressions for $a$ and $b$, \Ref{2.12}, entering the energy, are simply
\eq{3.2}{ a=\frac{1}{\ka},\quad b=\frac{1}{2\ka},
}	
resulting in \Ref{2.16} in the bound state energy
\eq{3.3}{ E_{bs} &= -\frac{\ep^4}{2\la\ka}=-\frac{(\ka^2-m^2)^2}{2\la\ka}.
}	
Note the approximation for small binding energies $\ep\simeq\sqrt{2m}\sqrt{\ka-m}$,
\eq{3.4}{ E_{bs} \raisebox{-4pt}{$\sim \atop \ka\to m$} -\frac{2m}{\la}(\ka-m)^2.
}	

An exact solution of the \rGP equation \Ref{1.12} without potential or matching condition is
\eq{3.5}{ u(x) &= \sqrt{\frac{2}{\la}}\frac{m}{\sqrt{2k^2-1}}
	\, ds\left(\frac{m (x+x_1)}{\sqrt{2k^2-1}},k\right),
}	
where $ds(z,k)$ is an elliptic Jacobi function and eq. \Ref{3.5} can be checked with eq. (22.13.23) in Ref. \cite{AbramowitzStegun2010}. There are two free parameters, the elliptic module $k$ and the shift $x_1$ of the coordinate. This function oscillates in $x$ unless $k=1$. Then, from table (22.5.4) in Ref. \cite{AbramowitzStegun2010}, we see that the solution is simply
\eq{3.6}{ u(x) &= \sqrt{\frac{2}{\la}}\ \frac{m}{\sinh(m(x+x_1))},
}	
which is decreasing in $x$.
Using the matching condition \Ref{2.3} instead of the potential and the symmetry under $x\to -x$, we modify the solution,
\eq{3.7}{ u(x) &= \sqrt{\frac{2}{\la}}\ \frac{m}{\sinh(m(|x|+x_1))},
}	
where the module $|x|$ is similar to that in \Ref{3.1}. We take this solution to be the solution of the \rGP equation,
\eq{3.8}{ \phi_0(x)= \sqrt{\frac{2}{\la}}\ \frac{m}{\sinh(m(|x|+x_1))},.
}	
Now inserting \Ref{3.8} into the matching condition \Ref{2.3} yields
\eq{3.9}{ x_1=\frac{1}{m}{\rm arctanh}\left(\frac{m}{\ka}\right),
}	
which completes an explicit solution of the \rGP equation with a delta function potential.

It should be noted that this solution is similar, but not identical, to a bright solution of the true (complex) \rGP equation.

The energy \Ref{2.5} of this solution can also be calculated explicitly,
\eq{3.10}{ E_{cond} &= -\frac{2(\ka+2m)(\ka-m)^2}{3\la}.
}	
The approximation for small $\ka-m$ yields exactly the same energy as the bound state, \Ref{3.4}, justifying the approximation \Ref{2.6} for the given example. Plots of the energies are shown in figure~\ref{fig:fig1}. It can be seen that the energy of the exact solutions is always lower than the energy of the approximation.

\begin{figure}[t]
	\includegraphics[width=10 cm]{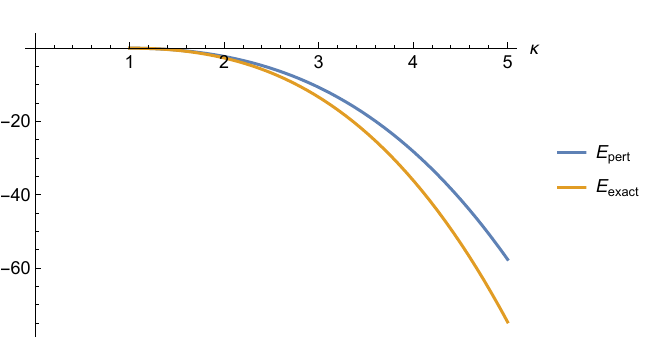}
	\caption{
		The energy \Ref{3.3} of the bound state solution and the energy \Ref{3.10} of the exact solution of the \rGP equation for the delta function potential \Ref{2.2} with $m=1$ and $\la=1$ as a function of the strengths $\ka$.}		\label{fig:fig1}	
\end{figure}

\section{\label{T4}Robin Boundary Condition}
In this section we consider the problem with the Robin boundary condition, \Ref{2.4}, on the left side of a finite interval $x\in [0,L]$. For sufficiently large parameter $\ka$ it leads to an unstable mode.  On the other side of the interval, we assume Dirichlet conditions for simplicity.
First we look at the bound state solution. It takes the form
\eq{4.1}{ \phi_{bs}(x) &= e^{-qx}-e^{q(x-2L)}, \quad q^2=\ep^2+m^2,
}
and it obeys the equation \Ref{1.6} with $V(x)=0$. The Dirichlet boundary condition in $x=L$ is satisfied by construction. From the Robin condition in $x=0$ we get the relation
\eq{4.2}{ q &= \ka \tanh(qL),
}
or, using \Ref{4.1}, expressed in terms of the bound state energy $\ep$,
\eq{4.3}{ \sqrt{\ep^2+m^2} &= \ka\tanh\left(\sqrt{\ep^2+m^2} L\right).
}
It has at most one solution. We are interested in a solution with real $\ep$. Its smallest value is $\ep=0$, which is only possible for $\ka L>1$. The expressions for the parameters $a$ and $b$, \Ref{2.12}, can be calculated explicitly, resulting in simple expressions. Substituting these into the energy \Ref{2.16}, we get
\eq{4.4}{E_{bs} &=- \frac{\left(2qL-\sinh(2qL)\right)^2}
	{2\la q(12qL-8\sinh(2qL)+\sinh(4ql))  } \,\ep^4,
}
for the bound state energy. Its dependence on $L$ is shown in the figure \ref{fig:fig7} { for $m=\lambda=1$, $\kappa=2$.}

As an approximation to the solution of the \rGP equation, we need this energy near the threshold at $\ep=0$. Here we have $q=m$. We denote the smallest $\ka$ by $\ka_c$ and from \Ref{4.2}
\eq{4.5}{ \ka_c&= \frac{m}{\tanh(mL)}
}
follows. We expand $q$ near the threshold, $q=m+\al(\ka-\ka_c)+\dots$, and from an expansion of the matching condition \Ref{4.2} we get
\eq{4.6}{ \al &= \frac{2\sinh^2(mL)}{\sinh(2mL)-2mL}.
}
Using \Ref{4.1}, this allows us to expand the energy, \Ref{2.16}, and we get
\eq{4.7}{ E_{bs}  \raisebox{-4pt}{$\sim \atop \ka\to \ka_c$}
	-\frac{8m\sinh^4(mL) }
	{\la\left( 12mL-8\sinh(2mL)+\sinh(4mL)\right)} (\ka-\ka_c)^2 .
}
If we also consider $L\to\infty$, we get with
\eq{4.8}{ E_{bs}  \raisebox{-4pt}{$\sim \atop \ka\to m,\ L\to\infty$}
	-\frac{m } 	{\la}  \, (\ka-m)^2,
}
which is similar to \Ref{3.4}.

For the example considered, an exact solution of the \rGP equation \Ref{1.12} can be obtained in the form
\eq{4.9}{ \phi_0(x) &=\sqrt{\frac{2}{\la}}\, \frac{ m k}{\sqrt{1+k^2}}\
	sc\left( \frac{ m (x-L)}{\sqrt{1+k^2}},\sqrt{1-k^2}\right),
}
where $sc(z,k)$ is also an elliptic Jacobian function.
This function, $\phi_0(x)$, satisfies the \rGP equation with $V(x)=0$ and, by construction, the Dirichlet boundary condition in $x=L$. The elliptic module $k$ can be taken in the interval $k\in[0,1]$. For values outside this interval, using the symmetry under $k\to 1/k$, one can use the relation $sc\left(z,\sqrt{1-k^2}\right)=\frac{1}{k}sc\left(k u,\sqrt{1-1/k^2}\right)$, which follows from a combination of (22.17.2) and (22.6(iv)) in Ref. \citen{AbramowitzStegun2010} and is not a new solution.

The Robin boundary condition \Ref{2.4} at $x=0$ results in the equation   the form
\eq{4.10}{ \ka\phi_0(0)+\phi'_0(0)=0
}
on the function $\Phi_0(x)$.

This is an equation for the elliptic module $k$. For the solutions one is left with numerical methods.

The energy $E_{cond}$, \Ref{2.5}, of the condensate can be calculated explicitly in terms of Jacobi's elliptic functions. The expression is generated by Mathematica and is too large to be shown in a paper.
The same applies to the expression for the Casimir force,
\eq{4.11}{ F_{cond} &= -\frac{\pa}{\pa L} E_{cond}.
}
Examples of energy are shown in figures \ref{fig:fig4a} (left panel) and \ref{fig:fig7} (left panel). A picture of the force is shown in figure \ref{fig:fig2} for $L>L_2$ (see eq. \Ref{4.14} below). This force is always repulsive.

Now we consider the solution near the threshold.
In the limiting case $k=0$, using $sn(z,0)=\sin(z)$, the solution \Ref{4.9} turns, up to an irrelevant factor, into \Ref{4.1} and one arrives just at the condition \Ref{4.5} of the bound state solution. By expanding \Ref{4.10} near $k=0$ in powers of $k$ on the one hand and in powers of $\ka-\ka_c$ on the other hand, the relation
\eq{4.12}{ \ka-\ka_c = \tilde\al k^2,\quad \tilde\al=-\frac{m(12mL-8\sinh[2mL]+\sinh(4mL))}{16\sinh^2(mL)}
}
follows in leading order. This relation allows to expand the energy in powers of $\ka-\ka_c$. You get exactly the same relation \Ref{4.7} as in the bound state approximation. Thus, also in the considered example, the bound state solution is a good approximation for the \rGP equation near the threshold.

To study the solutions of the equation \Ref{4.10}, we first consider the boundary values $k=0$ and $k=1$. As already mentioned, for $k=0$ this is the same condition \Ref{4.3} for $\ep=0$ as for the boundary solution. So the smallest $L$ is the same. We denote it by $L_0$ and note that
\eq{4.13}{ L_0=\frac{1}{m}{\rm arctanh}\left(\frac{m}{\ka}\right).
}
For k=1, using $sn(z,0)=\tan(z)$, we get from the boundary condition
\eq{4.14}{ L_1=\frac{1}{\sqrt{2}m}\arcsin\left(\frac{\sqrt{2}m}{\ka}\right),
	\quad L_2=\frac{1}{\sqrt{2}m}\left[\pi-\arcsin\left(\frac{\sqrt{2}m}{\ka}\right)\right],
}
(and solutions from higher branches of accsin, which we do not need). Now, if the matching condition \Ref{4.10} as a function of $k$ is continuous in the interval $k\in[0,1]$ (and it is), then it must have different signs at $k=0$ and $k=1$ to find a zero. This is indeed the case, as can be seen in Figure~\ref{fig:fig3} (left panel) for $L_0<L<L_1$. Accordingly, there is a solution to the matching condition, which we show in Figure~\ref{fig:fig3} (right panel).
In the plots we use $m=1$ and $\ka=2$, which with \Ref{4.13} and \Ref{4.14} results in $L_0=0.549306$ and $L_1=0.55536$. This interval is quite small.  The corresponding energy is shown in Figure~\ref{fig:fig4a} (left panel). The energy of the bound state solution, $E_{bs}$, is very close. The difference between the two is shown in Figure~\ref{fig:fig4a} (right panel). It is very small, with the condensate energy $E_{cond}$ being lower as expected.

\begin{figure}[h]
	\includegraphics[width=0.45\textwidth]{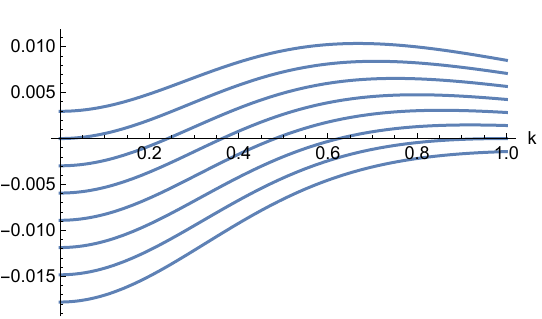}~~\includegraphics[width=0.45\textwidth]{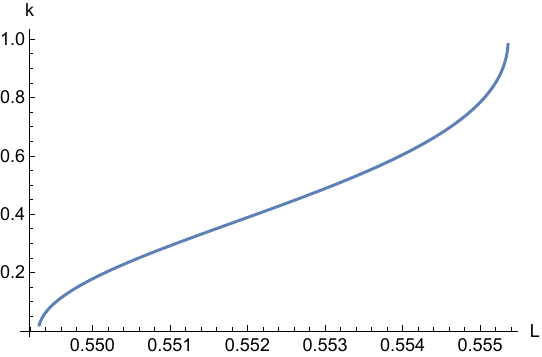}
	\caption{	The left panel shows the left-hand side of equation \Ref{4.10} as a function of $k$ for various $L$,	$	L_0 \lessapprox L \lessapprox L_1$ (from top to bottom). The top curve is for $L<L_0$ and the bottom curve is for $L>L_1$.  The right panel shows the solution $k$ of \Ref{4.10} for the region $L_0<L<L_1$.  The parameters are $m=1$, $\la=1$.}		\label{fig:fig3}	
\end{figure}

\begin{figure}[h]
	\includegraphics[width=0.45 \textwidth]{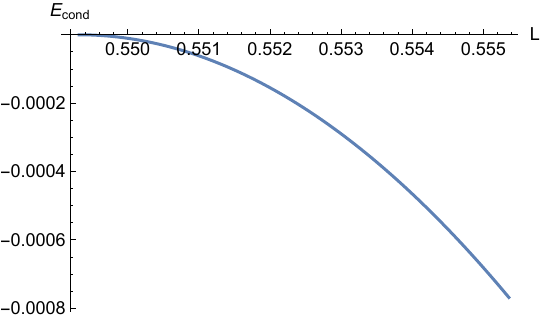}
	\includegraphics[width=0.45 \textwidth]{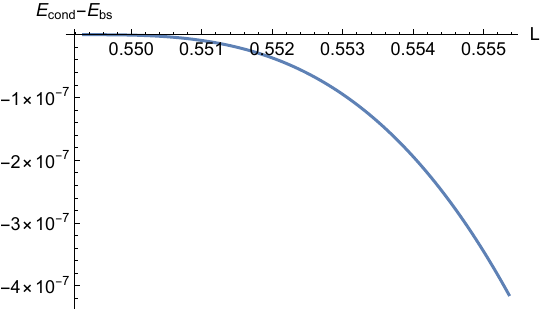}
	\caption{The energy $E_{cond}$ of the condensate in the region $L_0<L<L_1$ (left panel) and the difference to the perturbative energy, $E_{bs}$, (right panel).   }		\label{fig:fig4a}	
\end{figure}

The next interval is $L_1\le L\le L_2$, in the example we have $L_2=1.66608$. Examples are shown in figure~\ref{fig:fig5} (left panel). There are no solutions. The picture changes after $L_2$.  There are solutions, the corresponding $k$ are shown in figure~\ref{fig:fig5} (right panel).
The corresponding energy is shown in figure~\ref{fig:fig7} (left panel), together with the energy of the bound state.

\begin{figure}[h]
	\includegraphics[width=0.45\textwidth]{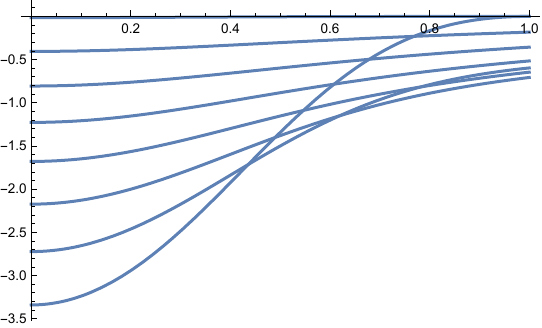}~~\includegraphics[width=0.45\textwidth]{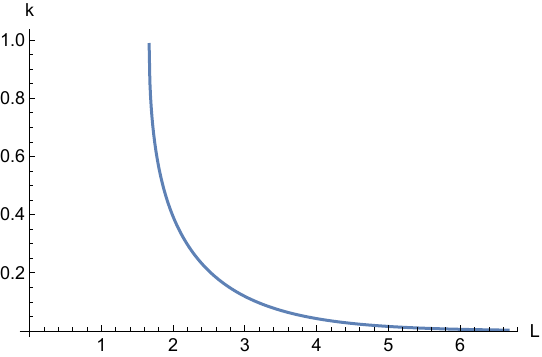}
	\caption{The left-hand side of  condition \Ref{4.10} as a function of $k$ for various $L$, from the interval $L_1<L<L_2$  (in order from top to bottom at the beginning of the curves). There is no solution.
		The solution in $k$ of \Ref{4.10} for the region $L>L_2$ is shown in the right panel.    }		\label{fig:fig5}	
\end{figure}

\begin{figure}[h]
	\includegraphics[width=0.47\textwidth]{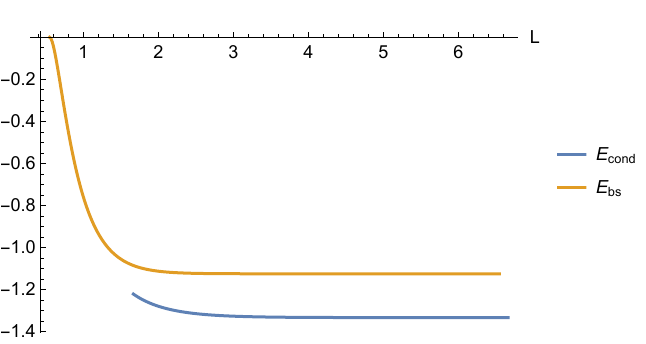}	\includegraphics[width=0.47\textwidth]{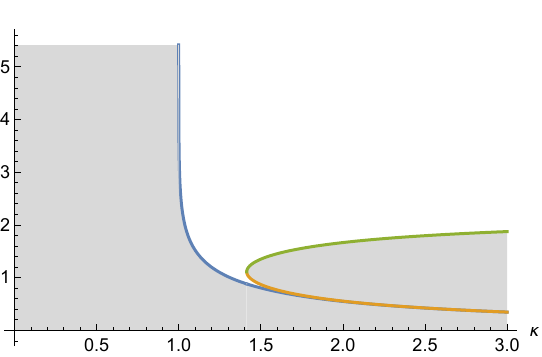}
	\caption{The energy $E_{bs}$ of the bound state (upper curve) and the energy $E_{cond}$ of the condensate (lower curve) are shown in the left panel.  The latter is shown in the region $L>L_2$, \Ref{4.14} { with $\kappa=2$, $m=1$}. For $L_0<L<L_1$ see figure \ref{fig:fig4a}. The right panel { shows the $L_i$, \Ref{4.13} and \Ref{4.14}, as a function of $\kappa$}, starting with $L_0$ as the lowest. In the shaded regions there are no solutions of the matching condition. }		\label{fig:fig7}	
\end{figure}

\begin{figure}[h]
	\includegraphics[width=0.45\textwidth]{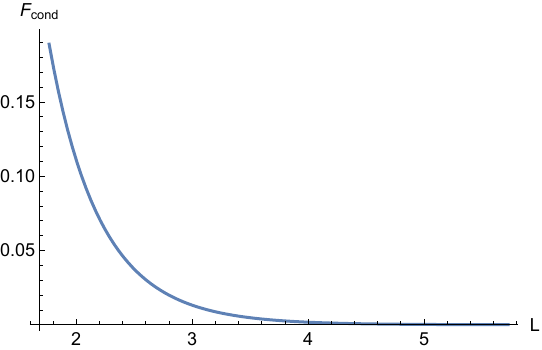}\includegraphics[width=0.45\textwidth]{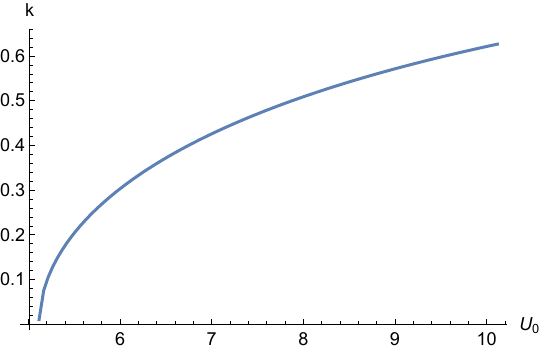}
	\caption{In the left panel, for the Robin boundary condition, the force \Ref{4.11} resulting from the condensate energy, for $m=1$, $\la=1$ and $\ka=2$ in the region $L>L_2$.
		In the right panel, for the potential hole, the elliptic module $k$ as a solution of the matching condition \Ref{5.6} as a function of the depth $U_0$ of the potential hole.	The parameters are $m=1$, $\la=1$ and $R=1$.	}
	\label{fig:fig2}	
\end{figure}

The regions in the ($\ka,L$)-plane are shown in figure \ref{fig:fig7} (right panel). There are no solutions in the shaded regions. For $\ka>\sqrt{2}m$ there is a gap between the solution regions ({\it k-gap}), i.e. a region with no solutions for the elliptic module $k$.  We attribute this to the nonlinearity of the \rGP equation. Above the k-gap, the matching condition \Ref{4.10} has more than one solution. However, the corresponding wave functions $\phi_0(x)$, \Ref{4.9}, have singularities in the region $0<x<L$ and are not physical. More details remain to be explored.
 
\section{\label{T5}Potential Hole}
The problem of a potential hole is one of the basic examples in a course on quantum mechanics, and here we repeat some well-known formulas. The bound state solution of equation \Ref{1.6} can be written in the form
\eq{5.1}{ \phi_{bs}(x) &= e^{-qR}\sin(px)\ \Theta(R-x)+\sin(pR)e^{-qx}\ \Theta(x-R),
}
where
\eq{5.2}{p&=\sqrt{U_0-m^2-\ep^2 },& q&=\sqrt{m^2+\ep^2}=\sqrt{U_0-p^2},
}
by inserting \Ref{5.1} into \Ref{1.6}. The boundary condition $\phi_{bs}(0)=0$ and the matching condition $\phi_{bs}(R-0)=\phi_{bs}(R+0)$ are satisfied by construction, and the continuity of the derivative leads to the matching condition
\eq{5.3}{ \tan(pR)=-\frac{p}{q}.
}
In this example, the free parameters are $U_0$ and $R$. Eq.\Ref{5.3} is a condition for the bound state energy $\ep$. The energy of the solution \Ref{5.1} can be calculated from \Ref{2.16} and is
\eq{5.4}{ E_{bs} &= -\frac{ \ep^4 U_0(1+qR)^2}
	{2\la q(3q^2(1+qR)+p^2(2+3qR))   }.
}

The exact solution of the \rGP equation \Ref{1.12} can be expressed in terms of an elliptic Jacobian function as
\eq{5.5}{ \phi_0(x) &=\sqrt{\frac{2}{\la}} \Big[
	\frac{\sqrt{U_0-m^2}\, k}{\sqrt{1+k^2} }
	\ sn\left(\frac{\sqrt{U_0-m^2}\, x}{\sqrt{1+k^2} },k\right)\ \Theta(R-x)
	\\\nn & ~~~~~~~~~~ +\frac{m}{\sinh(m(x+x_1))}\ \Theta(R-x)
	\Big].
}
The boundary condition $\phi_0(0)=0$ is satisfied by construction, for the remaining free parameters $k$ and $x_1$ we have two conditions, the continuity of the function and its derivative at $x=R$. By eliminating $x_1$, the remaining matching condition can be written in the form
\eq{5.6}{\phi'_{0}(R) &=-\phi_{0}(R)\sqrt{\frac{\la}{2}\phi_{0}(R)+m^2}.
}
The free parameters are $U_0$ and $R$, \Ref{5.6} is a condition for the elliptic module $k$. An example of the solution is shown in figure \ref{fig:fig2}, right panel.

The smallest $U_0$ and $R$ for which the system is critical follows from both conditions, \Ref{5.3} for $\ep=0$ and \Ref{5.6} for $k=0$, to satisfy the condition
\eq{5.7}{ \tan\left(\sqrt{U_0-m^2}R\right) =-\frac{\sqrt{U_0-m^2}}{m}
}

\begin{figure}[h]
	\includegraphics[width=0.47\textwidth]{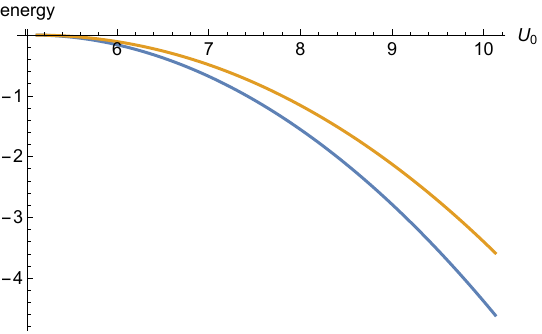}	
	\caption{
		The energy $E_{bs}$, \Ref{5.6}, of the bound state solution (upper curve) and the energy $E_{cond}$, \Ref{1.11}, of the exact solution (lower curve).
		In both cases the parameters are $m=1$, $\la=1$ and $R=1$.
	}		\label{fig:figY}	
\end{figure}

The energy of the approximate and the exact solution can be calculated explicitly. For the bound state solution we have
\Ref{5.4}. For the exact solution, a formula is generated by machine, which is too large to display in a paper. For both, plots can be easily calculated and we show an example in figure \ref{fig:figY}. Again, the exact energy is below the perturbative energy.

As in the previous examples, the energy can be expanded near the threshold \Ref{5.7}. We get the same expression,
\eq{5.8}{ E&= -\frac{m(m+R U_0)^2}{2\la U_0(m^2+2U_0+3mRU_0)}\,\Delta U_0+\dots,
}
where $U_0$ and $R$ are related by \Ref{5.6} and $\Delta U_0$ is a deviation from it,
for both the approximate and the exact solutions.
 
 \section{\label{T6}Fluctuations}
In this section we consider the quantum fluctuations of the field $\phi(x)$, i.e. its vacuum energy, for both cases, subcritical and critical. As an example, we consider the case with the Robin boundary condition, {studied} in section \ref{T4}. Since there is a finite interval, it makes sense to talk about the Casimir force.  We repeat some well known formulas that can be found in Refs. \citen{BKMM} and \citen{vass03-388-279}. The vacuum energy $E_0$ is defined as the half sum of the eigenfrequencies,
\eq{6.1}{ E_0&=\frac{1}{2}\sum_J (\om_J^2+m^2)^{\frac12-s}
	=\frac12\int_\ga\frac{d\om}{2\pi i}( \om^2+m^2)^{\frac12-s}\pa_\om \ln\Phi(\om),
}
of the kernel from the fluctuation part $\L_2$ of the Lagrangian in \Ref{1.10},
\eq{6.2}{\left(-\pa_x^2+{\mathcal V}(x)\right)\varphi_J(x) = \om_J^2 \varphi_J(x),
}
with ${\mathcal V}$ defined in \Ref{2.18}, ${\mathcal V}=V+3\lambda \phi^2$,  the appropriate boundary conditions are implied.
For convenience, in this section we have moved the mass parameter $m^2$ from the equation (see \Ref{1.6}) to the vacuum energy $E_0$.
The second expression in \Ref{6.1} is a representation of the sum by a contour integral, where the contour $\ga$ must contain all zeros of the function $\Phi(\om)$. This function, also called {\it mode generating function}, is defined to have the eigenfrequencies $\om_J$ as its zeros, including bound states with $\om_J=i\ep$. The parameter $s$ comes from the zeta-functional regularization. You need $\Re s>\frac12$ for convergence and $s\to0$ at the end. From \Ref{6.1}, deforming the contour, we get the representation
\eq{6.3}{ E_0 &= -\frac{\cos(\pi s)}{2\pi}\int_m^\infty d\xi \, (\xi^2-m^2)^{\frac12-s}\ \pa_\xi \ln\Phi(i\xi),
}
which is convenient to avoid the oscillations on the real $\om$-axis.

The divergent contributions in \Ref{6.3} can be obtained either from the heat kernel expansion or from the asymptotic expansion of $\Phi(i\xi)$ for $\xi\to\infty$. The (1+1)-dimensional analog of Eq. (4.30) in Ref.  \citen{BKMM} is
\eq{6.4}{E_0^{div}&= \frac{m^2}{8\pi}\left(\frac{1}{s}-\ln\frac{m^2}{4}-1\right)a_0
	+\frac{m}{4\sqrt{\pi}}\,a_{1/2}
	-\frac{1}{8\pi}\left(\frac{1}{s}-\ln\frac{m^2}{4}-2\right)\, a_1,
}	
where the heat kernel coefficients are taken from Ref. \citen{vass03-388-279},
\eq{6.5}{ a_0 &= L, & a_{1/2}&=0, & a_1 &= 2\kappa-\int_0^Ldx\,{\mathcal V}(x),
}
for the boundary conditions \Ref{2.4}. From the heat kernel expansion of the operator on the left side of \Ref{6.2}, the behavior of $\Phi(i\xi)$ for large $\xi$ immediately follows,
\eq{6.6}{ \ln \Phi(i\xi) &\simeq
	\sqrt{4\pi}a_0 \ \xi +\frac{a_{1/2}}{\sqrt{\pi}}\ln \xi-\frac{a_1}{2\xi}+\dots\,
	\equiv \ln\Phi^{as}+\dots\,.
}
{This expansion can also be obtained by an iteration like \Ref{2.9} of a corresponding integral equation.}

In the process of renormalization, the contribution of $\Phi^{as}$, i.e. the contribution $E_0^{div}$ of the corresponding (in our case up to $a_1$) first heat kernel coefficients, must be subtracted, interpreting this as a renormalization of the constants in the classical part. The renormalized vacuum energy is then
\eq{6.7}{ E_0^{ren}=E_0-E_0^{div},
}
where the limit $s\to 0$ must be taken. The pole terms cancel.
We will not go into further detail here and refer to Ref. \citen{BKMM}, Section 4.3.

With the above formulas we consider the example from section \ref{T4}, first for the subcritical case, then for the critical one.
In the latter case, we consider 2 versions. First, we use the exact condensate function \Ref{4.9}, and second, the approximate one, \Ref{4.1}.

\subsection{\label{T6.1}Robin condition in the subcritical case}
In this case the potential in eq. \Ref{6.2} is constant, ${\mathcal V}(x)=0$, and we have only the boundary conditions \Ref{2.4}. The solutions are well known and we take two independent solutions,
\eq{6.1.1}{ u_0(x)&=\cos(\om x)-\frac{\ka}{\om}\sin(\om x),&v_0(x)&=\frac{1}{\om}\sin(\om(x-L)),
}
and we assume $\ka<\ka_c$, \Ref{4.5}. The function $u_0(x)$ satisfies the boundary condition at $x=0$ and $v_0(x)$ satisfies the condition at $x=L$.
The Wronskian of these solutions is
\eq{6.1.2}{ W(\om) &=\cos(\om L)-\frac{\ka}{\om}\sin(\om L).
}
When the Wronskian is zero, the solutions are not independent and obey the constraints on both sides. Therefore, the Wronskian can be used for the mode generation function, $\Phi(\om)=2 W(\om)$.

For the vacuum energy \Ref{6.3} we need the expression on the imaginary axis,
\eq{6.1.4}{ \Phi(i\xi) &= 2\cosh(\xi L)-\frac{2\ka}{\xi}\sinh(\xi L).
}
Eq. \Ref{6.1.4} has a zero with real $\xi\equiv \ep$ (for $\ka>0$), which is the bound state. It is a solution of eq. \Ref{4.3} with the notation $\ep\to\xi$. In this subsection, since we are considering the subcritical case, we assume $\ep<m$ so that there is no zero on the $\xi$ axis in the integration domain in \Ref{6.3}.

To proceed, it is convenient to rewrite the logarithm of the mode generating function \Ref{6.1.4},
\eq{6.1.5}{\ln \Phi(i\xi) &=
	\xi L+\ln\left(1-\frac{\ka}{\xi}\right)+\ln\left(1+\frac{ \xi+\ka}{\xi-\ka}\, e^{-2\xi L}\right).
}
The first term is the Minkowski space contribution and must be dropped. The second does not depend on the distance $L$ and will not contribute to the Casimir force. The third contribution is convergent and provides the force.

From \Ref{6.1.5} also follows the asymptotic expansion for $\xi\to \infty$, which coincides with \Ref{6.4} and \Ref{6.5} and confirms the heat kernel coefficients \Ref{6.5} (we have ${\mathcal V}(x)=0$ in this subsection).

Defining
\eq{6.1.6}{\ln\Phi^{as} = -\frac{\ka}{\xi},
}
we write
\eq{6.1.7}{\ln\Phi(i\xi)-\ln\Phi^{as} =\ln\Phi_1+\ln\Phi_2
}
with
\eq{6.1.8}{\ln\Phi_1&= \ln\left(1-\frac{\ka}{\xi}\right)+\frac{\ka}{\xi},
	& \ln\Phi_2 &= \ln\left(1+\frac{\xi+\ka}{\xi-\ka}\, e^{-2\xi L}\right).
}
The expression in \Ref{6.1.7} decreases $\sim \xi^{-2}$ and the integral in \Ref{6.3} will converge.
According to \Ref{6.1.7} we also split the vacuum energy into two parts,
\eq{6.1.9}{E_0=E_0^{(1)}+E_0^{(2)}
}
with
\eq{6.1.10}{E_0^{(i)} &= -\frac{1}{2\pi} \int_m^\infty d\xi \ (\xi^2-m^2)^{\frac12}\, \pa_\xi \ln\Phi_i(i\xi),~~~~(i=1,2).
}
In this way the vacuum energy can be easily calculated numerically and the result is shown in figure \ref{fig:fluc}. The vacuum energy can have either sign, the force is repulsive, as known from previous work.
We mention Ref. \citen{rome02-35-12973}, where the same problem was solved for Robin conditions on both sides of the interval, resulting in an attractive force.

\begin{figure}[h]
	\includegraphics[width=0.47\textwidth]{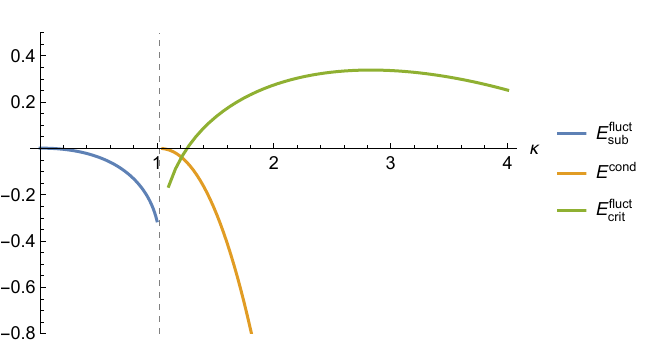}	\includegraphics[width=0.47\textwidth]{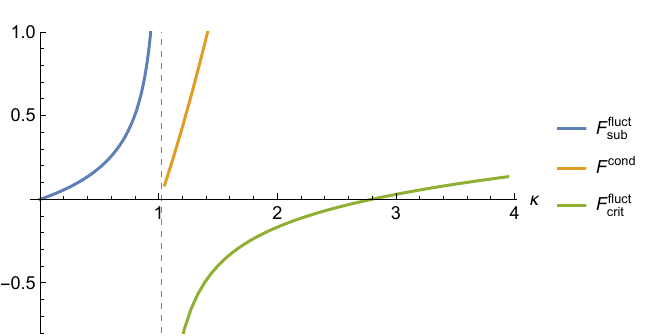}
	\includegraphics[width=0.47\textwidth]{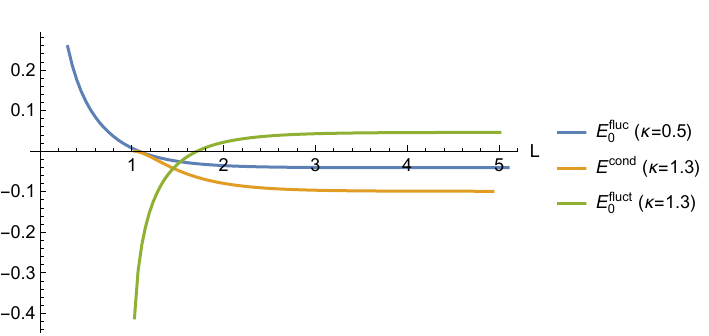}	\includegraphics[width=0.47\textwidth]{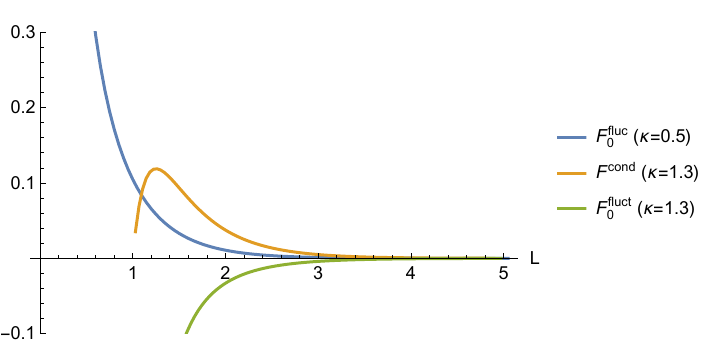}
	\caption{
		In the upper line, the vacuum energy (left panel) and the Casimir force (right panel) as a function of the strength $\ka$ of the Robin boundary condition for the separation $L=2.2$ with the critical value $\kappa_c(L)\simeq1.025$ (dashed vertical line). The lower line shows the corresponding quantities as a function of the separation $L$. In each case the vacuum energy for the subcritical case $\ka<m$, the vacuum energy and the condensate energy for the critical case ($\ka>m$) are shown. The parameters are $m=1$ and $\la=1$.
	}		\label{fig:fluc}	
\end{figure}

\subsection{\label{T6.2}Robin condition in the critical case with exact solution}
In this case, the equation for the fluctuations follows from $\L_2$ in \Ref{1.10} and reads
\eq{6.2.1}{ \left( - \pa_x^2+V_0(x) \right) \varphi_\om(x)=\om^2 \varphi_\om(x),
}
with
\eq{6.2.2}{ V_0(x)=3\la\phi_0(x)^2 ,
}
where $\phi_0(x)$ is the solution \Ref{4.9} of the \rGP equation. It must be complemented by the equation \Ref{4.10}. The fluctuations $\varphi_\om(x)$ must obey the boundary conditions \Ref{2.4} and the parameter $\kappa$ takes values that lead to an imaginary frequency solution in the case without condensate function.  By construction of the Lagrangian \Ref{1.9}, the equation \Ref{6.2.1} has no imaginary eigenvalues.

The calculation of the vacuum energy follows the same general lines as in the previous subsection, but now a solution of the equation \Ref{6.2.1} and with it the mode generating function can only be obtained numerically.
We use a scheme that is a direct generalization of the procedure in the previous subsection. Going directly to the imaginary frequency axis, we define two independent solutions, $u(v)$ and $v(x)$, with initial data derived from the boundary conditions,
\eq{6.2.3}{ \ka u(0)+u'(0)=0,~~~ v(L)=0.
}
These conditions leave us with 2 free constants. We take $u(0)=1$ and $v'(L)=\xi$. Numerically, the solution $u(x)$ can be integrated from the left ($x=0$) and the other, $v(x)$, from the right ($x=L$).
The mode generating function can again be constructed from the Wronskian
\eq{6.2.4}{W=u(x)v'(x)-u'(x)v(x)=\xi u(L),
}
where the last expression follows from the $x$-independence of $W$ and the boundary conditions.

The Wronskian $W$, \Ref{6.2.4}, turns out to be proportional to $e^{\xi L}$, which produces large numbers and is inconvenient for numerical evaluation. An easy way out is to separate this factor by introducing new functions,
\eq{6.2.5}{ u(x)=e^{\xi x}\tilde u(x),~~v(x)=e^{\xi(L-x)}\tilde v(x)
}
and to integrate the equations
\eq{6.2.6}{ \left(-\pa_x^2-2\xi\pa_x+V(x)\right)\tilde u(x)&=0,
	\\\nn \left(-\pa_x^2+2\xi\pa_x+V(x)\right)\tilde v(x)&=0.
}
The Wronskian becomes
\eq{6.2.7}{W=e^{\xi L}2\tilde u(L).
}
In fact, it is not necessary to calculate $\tilde v(x)$; it is sufficient to know $\tilde u(L)$.

For the vacuum energy, one needs to know the asymptotic expansion of $W$ for large $\xi$. It can be obtained by transforming the differential equations for $u(x)$ and $v(x)$,
\eq{6.2.8}{ \left( -\pa_x^2+\xi^2+{\mathcal V}(x)\right)\left\{{u(x)\atop v(x)}\right\} =0,
}
into integral equations and solve them by iteration. Using a Green's function obeying
\eq{6.2.9}{ \left(-\pa_x^2+\xi^2\right)G(x,x')=\delta(x-x'),
}
which read
\eq{6.2.10}{ u(x)&=u_0(x)-\int_0^L dx'G(x,x'){\mathcal V}(x')u(x'),
	\\\nn v(x)&=v_0(x)-\int_0^L dx'G(x,x'){\mathcal V}(x')v(x').
}
For the initial function, defined with $V(x)=0$, we take \Ref{6.1.1}.
In terms of these functions, the Green's function is
\eq{6.2.12}{ G(x,x')=\frac{u_0(x_<)v_0(x_>)}{W_0},~~~W_0=u_0(x)v_0'(x)-u_0'(x)v_0(x),
}
and $W_0$ coincides with \Ref{6.1.4}. The iteration starts by inserting the initial functions on the right side. One step is enough for our purposes, and for large $\xi$ you get
\eq{6.2.13}{W= e^{\xi L}\left(1-\frac{\ka}{\xi}\right)
	\left(1+\frac{1}{2\xi}\int_0^Ldx \, {\mathcal  V}(x)\right)+\dots\,.
}
Taking this Wronskian as the mode generating function, a comparison with \Ref{6.6} for large $\xi$ confirms $a_1$ in \Ref{6.5}.

Now we take the mode generating function without the exponential term from \Ref{6.2.7} and its asymptotic expansion from \Ref{6.2.13} is
\eq{6.2.14}{ \Phi(i \xi)=2\tilde u(L),~~
	\Phi(i\xi)\simeq\left(1-\frac{1}{2\xi}\left(2\ka-\int_0^Ldx \, {\mathcal V}(x)\right)\right)+\dots\,.
}
In the given approximation we define
\eq{6.2.15}{\ln\Phi^{as}=\frac{-1}{2\xi} \left(2\ka-\int_0^Ldx \, {\mathcal V}(x)\right),
}
which corresponds to the general formula \Ref{6.6}.
Inserting this expression in \Ref{6.7}, using \Ref{6.3}, we get for the renormalized vacuum energy
\eq{6.2.16}{ E_0^{ren}=-\frac{1}{2\pi}\int_m^\infty d\xi\, (\xi^2-m^2)^{\frac12}\,
	\pa_\xi \left( \ln \Phi(i\xi)-\ln\Phi^{as}  	\right).
}
The expression is finite by construction. For numerical integration, it is convenient to integrate by parts to get rid of the derivative,
\eq{6.2.17}{ E_0^{ren}=\frac{1}{2\pi}\int_m^\infty d\xi\, \frac{\xi}{ (\xi^2-m^2)^{\frac12}}\,
	\left( \ln \Phi(i\xi)-\ln\Phi^{as}  	\right).
}
Then the vacuum energy can be calculated. An example is shown in figure \ref{fig:fluc}. This energy is negative and the corresponding force is repulsive.

\subsection{\label{T6.3}Robin condition in the critical case with approximate solution}
In this subsection we calculate the vacuum energy in the critical case with the potential ${\mathcal V}(x)$ taken from the approximate solution $\phi_{bs}(x)$, \Ref{4.1} instead of $\phi_0(x)$, \Ref{4.9}. The potential in the equation \Ref{6.2.1} for the fluctuations is
\eq{6.3.1}{{\mathcal V}(x)=3\la \phi_{bs}(x)^2.
}
Both potentials are compared in the figure \ref{fig:fluc2} for $L=3$.

With this potential, \Ref{6.3.1}, the equation \Ref{6.2.1} for the fluctuations has modified Mathieu functions as solution. However, their implementation in Mathematica is not reliable. Therefore, for the Wronskian $W$, \Ref{6.1.2}, we used the same numerical procedure as in the previous subsection.

The results of the calculation of the vacuum energy are shown in figure \ref{fig:fluc2} for both cases, the exact solution and the approximate one.  As can be seen, there is only a very small difference despite the rather large difference between the corresponding potentials shown in the left panel of figure \ref{fig:fluc2}. The reason is probably that most of the vacuum energy comes from the boundary conditions, which are of course the same in both cases.

\begin{figure}[h]
	\includegraphics[width=0.47\textwidth]{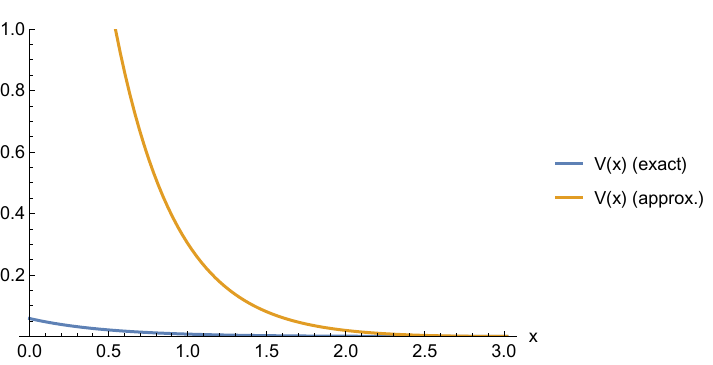}	\includegraphics[width=0.47\textwidth]{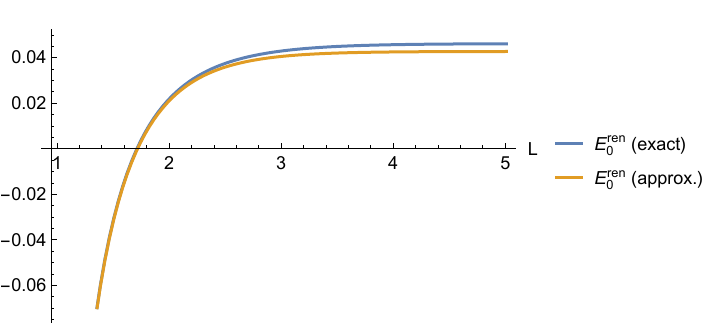}
	
	\caption{
		In the left panel, the background potentials $V(x)$ for the exact solution $\phi_0(x)$ of the \rGP equation as a function in the interval with $L=3$.
		The right panel shows the vacuum energies calculated with the two background potentials shown in the left panel.
		The parameters are $m=1$, $\la=1$, and $\ka=1.3$.
	}		\label{fig:fluc2}	
\end{figure}

\section{\label{T7}Conclusions}

We have studied the vacuum energy in systems with unstable modes, i.e. those with bound states with imaginary frequencies. Since there is a threshold for their existence, we have studied vacuum fluctuations in both subcritical and critical regimes.

A (1+1) system with self-interaction and static background potential was considered. In a situation with unstable modes, a condensate is generated and one must split the field into a classical static part (condensate) and the quantum field. In all cases studied in this paper, the nonlinear equation defining the condensate (\rGP equation) can be solved exactly in terms of elliptic Jacobi functions. This is an exceptional feature of the one-dimensional case and does not hold for all potentials $V(x)$. In general, one is left with approximate or numerical methods. An approximate method is a kind of mean-field approximation where the nonlinearity is replaced by a constant, see Eq. \Ref{2.5a}. Using the exact solutions, we show that this approximation works well for weak criticality. However, in the example with the Robin boundary condition, we observed a parameter region ('k-gap') with no solution. Of course, the approximate solution is insensitive to the k-gap and gives a 'false positive' answer here.

Our main interest is the influence of the condensate on the Casimir effect, which makes sense in the second example with a finite spatial interval. We use the known methods. Since there are no exact solutions for the fluctuations of the considered background potentials, we were left with numerical methods, which are quite simple in this case.

{By construction, the spectrum of these fluctuations is free of unstable modes, i.e. those with imaginary frequencies corresponding to bound states, below the mass. For the exact solution, this is guaranteed "by construction" by solving the \rGP equation. For the approximate solution, which has an energy higher than the true vacuum, there is no guarantee. However, in our numerical examples we did not observe any such problem.

In the subcritical region, the Casimir force is repulsive, as expected from a configuration with two different boundaries. In the case of a second Robin condition (instead of the Dirichlet condition), the force would be attractive. In the critical region the situation is different. Here we have two contributions, one from the condensate (whose energy also depends on the separation between the boundaries) and the other from the vacuum fluctuations. Both are of similar strength, changing their relative weight depending on the parameters. The results are shown in figure \ref{fig:fluc}. It is interesting to note that the condensate always contributes a repulsive force. In the case of equal boundary conditions on both sides, the vacuum force would be attractive, but could be compensated by the repulsive condensate force.

Finally, we compared the vacuum energies for the background resulting from the exact solution of the \rGP equation and from the approximate solution. These potentials are compared in the left panel of figure \ref{fig:fluc2}. As can be seen, there is quite a large difference between them. However, the corresponding vacuum energies, shown in the right panel of this figure, are very close. This means that the influence of the background potential on the vacuum energy is quite small compared to the influence of the boundary conditions.

An interesting topic for further investigation is the k-gap observed in section \ref{T4}. While bifurcations are known to result from nonlinearity, such a gap has not been observed to our knowledge. Of course, it is also interesting to consider the Casimir effect for more realistic condensate models.

\section*{Acknowledgements}
We gratefully acknowledge valuable and helpful discussions with D.N. Voskre\-sensky.

 \bibliographystyle{unsrt}
\bibliography{C:/Users/Bordag/WORK/Literatur/bib/papers}

\end{document}